\documentclass[11pt,preprintnumbers,titlepage,nofootinbib]{revtex4-2}
\usepackage[latin9]{inputenc}
\usepackage{amsmath}
\usepackage{amssymb}
\usepackage{graphicx}
\usepackage[unicode=true,
 bookmarks=false,
 breaklinks=false,pdfborder={0 0 1},backref=section,colorlinks=false]
 {hyperref}
\hypersetup{
 colorlinks,linkcolor=blue,citecolor=blue,urlcolor=blue}

\makeatletter

\providecommand{\tabularnewline}{\\}
\newcommand{\lyxdot}{.}

\usepackage{textcomp}
\usepackage{wasysym}
\usepackage{array}
\usepackage{multirow}
\usepackage{wasysym}
\usepackage{wasysym}
\usepackage{amsfonts}
\usepackage{subfigure}
\usepackage{cleveref}
\usepackage{listings}
\usepackage{xcolor}
\usepackage{array}
\usepackage{url}
\usepackage{color}
\usepackage{subfigure}

\providecommand{\tabularnewline}{\\}
\DeclareFontEncoding{LGR}{}{}

\ProvideTextCommand{\~}{LGR}[1]{\char126#1}
\DeclareTextSymbolDefault{\textquotedbl}{T1}

\@ifundefined{textcolor}{}{
\definecolor{BLACK}{gray}{0}
\definecolor{WHITE}{gray}{1}
\definecolor{RED}{rgb}{1,0,0}
\definecolor{GREEN}{rgb}{0,1,0}
\definecolor{BLUE}{rgb}{0,0,1}
\definecolor{CYAN}{cmyk}{1,0,0,0}
\definecolor{MAGENTA}{cmyk}{0,1,0,0}
\definecolor{YELLOW}{cmyk}{0,0,1,0}
\definecolor{ballblue}{rgb}{0.13, 0.67, 0.8}
\definecolor{bleudefrance}{rgb}{0.19, 0.55, 0.91}
\definecolor{blue(ncs)}{rgb}{0.0, 0.53, 0.74}
\definecolor{darkpastelgreen}{rgb}{0.01, 0.75, 0.24}
\definecolor{darkspringgreen}{rgb}{0.09, 0.45, 0.27}
\definecolor{denim}{rgb}{0.08, 0.38, 0.74}
\definecolor{electricviolet}{rgb}{0.56, 0.0, 1.0}
}

\makeatother

\begin{document}
\preprint{CTP-SCU/2024008}
\title{Observational Signatures of Traversable Wormholes }
\author{Yiqian Chen$^{a,b}$}
\email{chenyiqian@ucas.ac.cn}

\author{Lang Cheng$^{a}$}
\email{chenglang@stu.scu.edu.cn}

\author{Peng Wang$^{a}$}
\email{pengw@scu.edu.cn}

\author{Haitang Yang$^{a}$}
\email{hyanga@scu.edu.cn}

\affiliation{$^{a}$Center for Theoretical Physics, College of Physics, Sichuan
University, Chengdu, 610064, China}
\affiliation{$^{b}$School of Fundamental Physics and Mathematical Sciences, Hangzhou
Institute for Advanced Study, University of Chinese Academy of Sciences,
Hangzhou, 310024, China}
\begin{abstract}
In this paper, we study the observational signatures of traversable
Simpson-Visser wormholes illuminated by luminous celestial spheres
and orbiting hot spots. We demonstrate that when light sources and
observers are on the same side of the wormholes, the images of the
wormholes mimic those of black holes. However, when the light sources
are positioned on the opposite side from observers, photons traversing
the wormhole throat generate distinct observational signatures. Specifically,
unlike black hole images, the wormhole images are confined within
the critical curve, resulting in smaller centroid variations. Furthermore,
the light curve of hot spots can exhibit additional peaks. 
\end{abstract}
\maketitle
\tableofcontents{}

{}

\section{Introduction}

The groundbreaking images of the supermassive black holes candidates
M87{*} and Sgr A{*}, captured by the Event Horizon Telescope (EHT)
collaboration, have opened new avenues for understanding the nature
in the strong field regime \cite{EventHorizonTelescope:2019dse,EventHorizonTelescope:2019uob,EventHorizonTelescope:2019jan,EventHorizonTelescope:2019ths,EventHorizonTelescope:2019pgp,EventHorizonTelescope:2019ggy,EventHorizonTelescope:2021bee,EventHorizonTelescope:2021srq,EventHorizonTelescope:2022xnr,EventHorizonTelescope:2022vjs,EventHorizonTelescope:2022wok,EventHorizonTelescope:2022exc,EventHorizonTelescope:2022urf,EventHorizonTelescope:2022xqj}.
These images reveal a characteristic feature: a dark interior region
surrounded by a bright ring, which is in good agreement with theoretical
predictions for Kerr black holes. This feature emerges from the strong
gravitational lensing of light near light rings (or photon spheres
in spherically symmetric black holes) \cite{Synge:1966okc,Bardeen:1972fi,Bardeen:1973tla,Virbhadra:1999nm,Claudel:2000yi,Virbhadra:2008ws,Bozza:2009yw,Virbhadra:2022iiy}.
As a result, black hole images encode valuable information of the
black hole geometry and have spurred extensive research \cite{Amarilla:2010zq,Amarilla:2011fx,Wei:2013kza,Atamurotov:2015xfa,Dastan:2016vhb,Wang:2017hjl,Ayzenberg:2018jip,Stuchlik:2019uvf,Ma:2019ybz,Guo:2019lur,Zhu:2019ura,Ma:2020dhv,Hu:2020usx,Kruglov:2020tes,Wei:2020ght,Zeng:2020dco,Guo:2020zmf,Zhang:2020xub,Zhong:2021mty,Addazi:2021pty,He:2022opa}.

However, the finite resolution of the EHT observations allows alternative
explanations beyond black holes. Certain horizonless Ultra-Compact
Objects (UCOs) can exhibit light rings (or photon spheres) similar
to black holes, mimicking their behavior in observational simulations
\cite{Cunha:2017qtt,Narzilloev:2020peq,Guo:2020qwk,Herdeiro:2021lwl}.
Therefore, distinguishing these UCOs from black holes is a crucial
topic. For example, various studies have proposed echo signals in
late-time waveforms as a potential discriminant, arising from the
presence of a reflective surface or an extra photon sphere in specific
UCO models \cite{Cardoso:2016rao,Mark:2017dnq,Bueno:2017hyj,Konoplya:2018yrp,Wang:2018cum,Wang:2018mlp,Cardoso:2019rvt,GalvezGhersi:2019lag,Liu:2020qia,Yang:2021cvh,Ou:2021efv}.
Moreover, the general relativistic Poynting-Robertson effect and epicyclic frequencies can be employed as tools to investigate the potential existence of wormholes \cite{DeFalco:2020afv,DeFalco:2021btn}.
Additionally, asymmetric thin-shell wormholes with two photon spheres
have been found to exhibit double shadows and an extra photon ring
in their images \cite{Wang:2020emr,Wielgus:2020uqz,Guerrero:2021pxt,Peng:2021osd,Guerrero:2022qkh}.
However, the existence of multiple photon spheres outside the event
horizon has also been reported for a class of hairy black holes within
certain parameter spaces \cite{Herdeiro:2018wub,Wang:2020ohb,Gan:2021pwu,Guo:2021zed,Guo:2021ere}.
These multiple photon spheres in black holes can introduce features
similar to those observed in wormholes, including echo signals \cite{Guo:2022umh},
double shadows \cite{Guo:2021ere} and extra photon rings \cite{Gan:2021xdl,Guo:2022muy}.
These findings highlight the ongoing challenge of differentiating
between wormholes and black holes, necessitating the development of
further discriminatory methods for UCOs and black holes.

Recent observations of flaring activity near black holes, particularly
the recurrent detections close to Sgr A{*} \cite{Witzel:2020yrp,Michail:2021pgd,GRAVITY:2021hxs},
have garnered significant attention. While the underlying mechanism
remains unclear, it is generally believed to be attributed to magnetic
reconnection within magnetized accretion disks \cite{Dexter:2020cuv,Scepi:2021xgs,ElMellah:2021tjo}.
Nevertheless, orbital hot spots have been employed to understand the
observational signatures of these flares \cite{abuter2018detection,GRAVITY:2020lpa,Wielgus:2022heh,Huang:2024wpj}.
Research on the observational signatures of hot spots has been extended
to hairy black holes \cite{Chen:2024ilc} and various UCOs \cite{Li:2014coa,Rosa:2022toh,Chen:2024}
to differentiate between black holes and UCOs.

This paper investigates a static, spherically symmetric regular spacetime
proposed by Simpson and Visser \cite{Simpson:2018tsi}. This spacetime
interpolates between black holes, black-bounces and wormholes through
a parameter denoted by $a$. Previous studies have shown that for
$0<a<3M$, the regular spacetime exhibits the same shadow as a Schwarzschild
black hole with identical mass and distance \cite{Nascimento:2020ime,Lima:2021las,Bronnikov:2021liv}.
Notably, based on observations of the M87 galaxy's center by the EHT,
the parameter is estimated to be $a\approx4.2M$ \cite{Tsukamoto:2020bjm}. Simultaneously,
a rotating generalization of the Simpson-Visser metric within
a specific region of the parameter space can effectively model the observed
size and deviation from circularity of M87{*}'s shadow \cite{Shaikh:2021yux}.
These imply that the wormhole can closely mimic the observational
behavior of black holes. However, prior research has solely considered
scenarios where the light source and observer reside on the same side
of the wormhole. It is both natural and necessary to investigate scenarios
where the light source and the observer are located on different sides
of the wormhole. Intuitively, such a scenario would yield significantly
different observational results, potentially aiding in the distinction
between wormholes and black holes.

Our work focuses on the optical appearances of traversable Simpson-Visser
wormholes illuminated by luminous celestial spheres and the hot spots.
In Section \ref{sec:Set-up}, we review the Simpson-Visser spacetime
and discuss circular orbits for both massless and massive particles.
Sections \ref{sec:Celestial-Sphere} and \ref{sec:Hot-Spot} present
our numerical simulations for the celestial sphere and the hot spot,
respectively. Finally, we summarize our key findings in Section \ref{sec:Conclusions}.
Throughout the paper, we adopt the convention $G=c=1$ .

\section{Set up}

\label{sec:Set-up}

The Simpson-Visser spacetime can be described by the line element
presented in \cite{Simpson:2018tsi}, 
\begin{align}
ds^{2} & =-f\left(r\right)dt^{2}+\frac{1}{f\left(r\right)}dr^{2}+\left(r^{2}+a^{2}\right)\left(d\theta^{2}+\sin^{2}\theta d\phi^{2}\right),\nonumber \\
f\left(r\right) & =1-\frac{2M}{\sqrt{r^{2}+a^{2}}},
\end{align}
where $M$ is the ADM mass, and $a$ is a parameter to regularize
the spacetime. The spacetime can transition from a Schwarzschild black
hole to a traversable wormhole by increasing $a$ from $0$. Specifically,
the cases with $a=0$ and $a>2M$ correspond to a Schwarzschild black
hole and a traversable wormhole, respectively. Additionally, intermediate
states exist, including a black-bounce for $0<a<2M$ and a one-way
wormhole for $a=2M$. In this paper, we focus on traversable wormholes
with $a>2M$. Note that traversable wormholes possess two distinct
spacetimes, described by $r>0$ and $r<0$, respectively. These two
spacetimes are connected at the throat, located at $r=0$.

The motion of test particles in the spacetime is governed by the geodesic
equations, 
\begin{equation}
\frac{dx^{\mu}}{d\lambda}=p^{\mu},\quad\frac{dp^{\mu}}{d\lambda}=-\Gamma_{\rho\sigma}^{\mu}p^{\rho}p^{\sigma},\label{eq:geoeqn}
\end{equation}
where $\lambda$ is an affine parameter, and $\Gamma_{\rho\sigma}^{\mu}$
is the Christoffel symbol. Due to the spherically symmetric and static
nature of the spacetime, three conserved quantities arise that characterize
the geodesics, 
\begin{equation}
E=-p_{t},\quad L_{z}=p_{\phi},\quad L^{2}=p_{\theta}^{2}+L_{z}^{2}\csc^{2}\theta.\label{eq:EL}
\end{equation}
For a massless particle, $E$, $L_{z}$ and $L$ represent the total
energy, the angular momentum along the symmetry axis, and the total
angular momentum, respectively. For a massive particle, these quantities
describe the corresponding values per unit mass. Besides, the Hamiltonian
constraint, $\mathcal{H}=g_{\mu\nu}p^{\mu}p^{\nu}/2=\epsilon/2$,
introduces a fourth constant. This constant takes the value $\epsilon=0$
and $-1$ for massless and massive particles, respectively. In spherically
symmetric spacetimes, the trajectories of test particles are uniquely
characterized by the impact parameter $b\equiv L/E$ and the constant
$\epsilon$. According to the Hamiltonian constraint and eqns. $\left(\ref{eq:geoeqn}\right)$
and $\left(\ref{eq:EL}\right)$, the radical geodesic equation can
be expressed as 
\begin{equation}
\left(\frac{dr}{d\lambda}\right)^{2}+V_{\text{eff}}\left(r\right)=b^{-2},
\end{equation}
where the effective potential is defined as 
\begin{equation}
V_{\text{eff}}\left(r\right)=\left(-\frac{\epsilon}{L^{2}}+\frac{1}{r^{2}+a^{2}}\right)\left(1-\frac{2M}{\sqrt{r^{2}+a^{2}}}\right).
\end{equation}

\begin{figure}[ptb]
\includegraphics[width=0.9\textwidth]{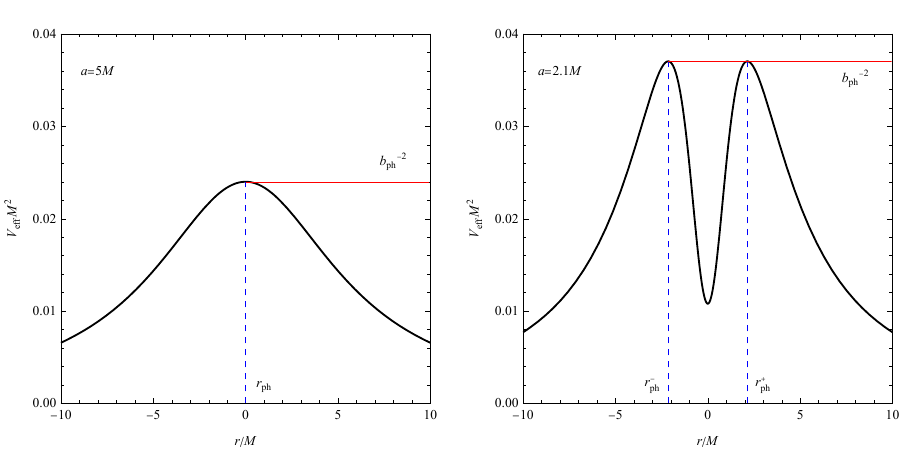}\caption{The effective potentials for photons. \textbf{Left Panel:} The traversable
wormhole with $a=5M$ possesses a single photon sphere located at
$r=0$, corresponding to the critical impact parameter $b_{\text{ph}}=5\sqrt{5/3}M$.
\textbf{Right Panel: }The traversable wormhole with $a=2.1M$ possesses
two photon spheres located at $r_{\text{ph}}^{+}=3\sqrt{51}M/10$
and $r_{\text{ph}}^{-}=-3\sqrt{51}M/10$, corresponding to the same
critical impact parameter $b_{\text{ph}}=3\sqrt{3}M$. }
\label{FIG: veff} 
\end{figure}

In the case of photons, we consider unstable circular orbits of radius
$r_{\text{ph}}$, which correspond to the maximum of the effective
potential, 
\begin{equation}
V_{\text{eff}}\left(r_{\text{ph}}\right)=b_{\text{ph}}^{-2},\quad V_{\text{eff}}^{\prime}\left(r_{\text{ph}}\right)=0,\quad V_{\text{eff}}^{\prime\prime}\left(r_{\text{ph}}\right)<0.
\end{equation}
These unstable circular orbits form a photon sphere at $r_{\text{ph}}$,
which has a significant impact on observations. Depending on the value
of $a$, the wormhole may possess either one or two photon spheres.
Specifically, when $a\leq3M$, the wormhole has a single photon sphere
at $r_{\text{ph}}=0$, and when $2M<a<3M$, the wormhole has two photon
spheres at $r_{\text{ph}}^{\pm}=\pm\sqrt{9M^{2}-a^{2}}$. To illustrate
the effects of different numbers of photon spheres on observational
phenomena, we will take the wormhole of $Q=5M$ with a single photon
sphere and the wormhole of $Q=2.1M$ with two photon spheres as examples.
The effective potentials for these two cases are shown in FIG. \ref{FIG: veff}.

For massive particles, their stable circular orbits can exist within
a certain region. The inner edge of this region corresponds to an
innermost stable circular orbit (ISCO), satisfying the following conditions
\begin{equation}
V_{\text{eff}}\left(r_{e}\right)=b_{e}^{-2},\quad V_{\text{eff}}^{\prime}\left(r_{e}\right)=0,\quad V_{\text{eff}}^{\prime\prime}\left(r_{e}\right)=0.
\end{equation}
When $2M<a<6M$, the ISCOs are located at $r_{e}^{\pm}=\pm\sqrt{36M^{2}-a^{2}}$.
If we consider a hot spot orbiting the wormhole at the ISCO on the
equatorial plane, its energy and angular momentum per unit mass are
given by $E_{e}=\sqrt{8/15}M$ and $L_{e}=6M/\sqrt{5}$, respectively.
Consequently, the corresponding angular velocity and period are $\Omega_{e}=M^{-1}/6\sqrt{6}$
and $T_{e}=12\sqrt{6}\pi M$, respectively.

To obtain the images of some light sources, we use the backward ray-tracing
method to numerically compute light rays from observers to light sources.
In the local frame of a static observer at $\left(t_{o},r_{o},\theta_{o},\phi_{o}\right)$,
the photon's local 4-momentum $p^{(\mu)}$ can be expressed by the
initial 4-momentum $p_{o}^{\mu}$, 
\begin{equation}
p^{\left(t\right)}=\frac{p_{o}^{t}}{f\left(r_{o}\right)},\quad p^{\left(r\right)}=f\left(r_{o}\right)p_{o}^{r},\quad p^{\left(\theta\right)}=\sqrt{r_{o}^{2}+a^{2}}p_{o}^{\theta},\quad p^{\left(\phi\right)}=\sqrt{r_{o}^{2}+a^{2}}\left\vert \sin\theta_{o}\right\vert p_{o}^{\phi}.\label{eq:p-local}
\end{equation}
Furthermore, considering the observation angles $\alpha$ and $\beta$
as defined in \cite{Cunha:2016bpi}, the components of the local 4-momentum
become 
\begin{equation}
p^{\left(r\right)}=p^{\left(t\right)}\cos\alpha\cos\beta,\quad p^{\left(\theta\right)}=p^{\left(t\right)}\sin\alpha,\quad p^{\left(\phi\right)}=p^{\left(t\right)}\cos\alpha\sin\beta.\label{eq:observation angles}
\end{equation}
These relationships $\left(\ref{eq:p-local}\right)$ and $\left(\ref{eq:observation angles}\right)$
connect the initial conditions for light rays to the observation angles.
In the image plane, we define the Cartesian coordinates as 
\begin{equation}
x\equiv-r_{o}\beta,\quad y\equiv r_{o}\alpha.
\end{equation}

\section{Celestial Sphere}

\label{sec:Celestial-Sphere}

This section investigates observations of traversable wormholes illuminated
by a celestial sphere. This model simulates the optical appearance
of the universe as lensed by the wormholes. To illustrate the image
of the celestial sphere, we position a luminous celestial sphere at
$r_{\text{cs}}=50M$ or $-50M$, while a static observer is situated
at $r_{o}=10M$, $\theta_{o}=\pi/2$ and $\phi_{o}=\pi$. The celestial
sphere is divided into four quadrants (colored green, red, blue and
yellow) corresponding to the upper left, upper right, lower left and
lower right regions relative to the observer. Additionally, a grid
of black lines is overlaid, representing lines of constant longitude
and latitude, where adjacent lines are separated by $\pi/18$. For
a more detailed discussion on the external view of the celestial sphere,
refer to {\cite{Bohn:2014xxa,Cunha:2015yba,Chen:2023trn}}. To generate
a simulated image, we vary the observation angles and numerically
integrate the trajectories of $2000\times2000$ photons until they
intersect with the celestial sphere or reach the cutoff radius at
$\left\vert r\right\vert =$ $50M$.

\begin{figure}[ptb]
\begin{tabular}{cc}
\includegraphics[width=0.4\textwidth]{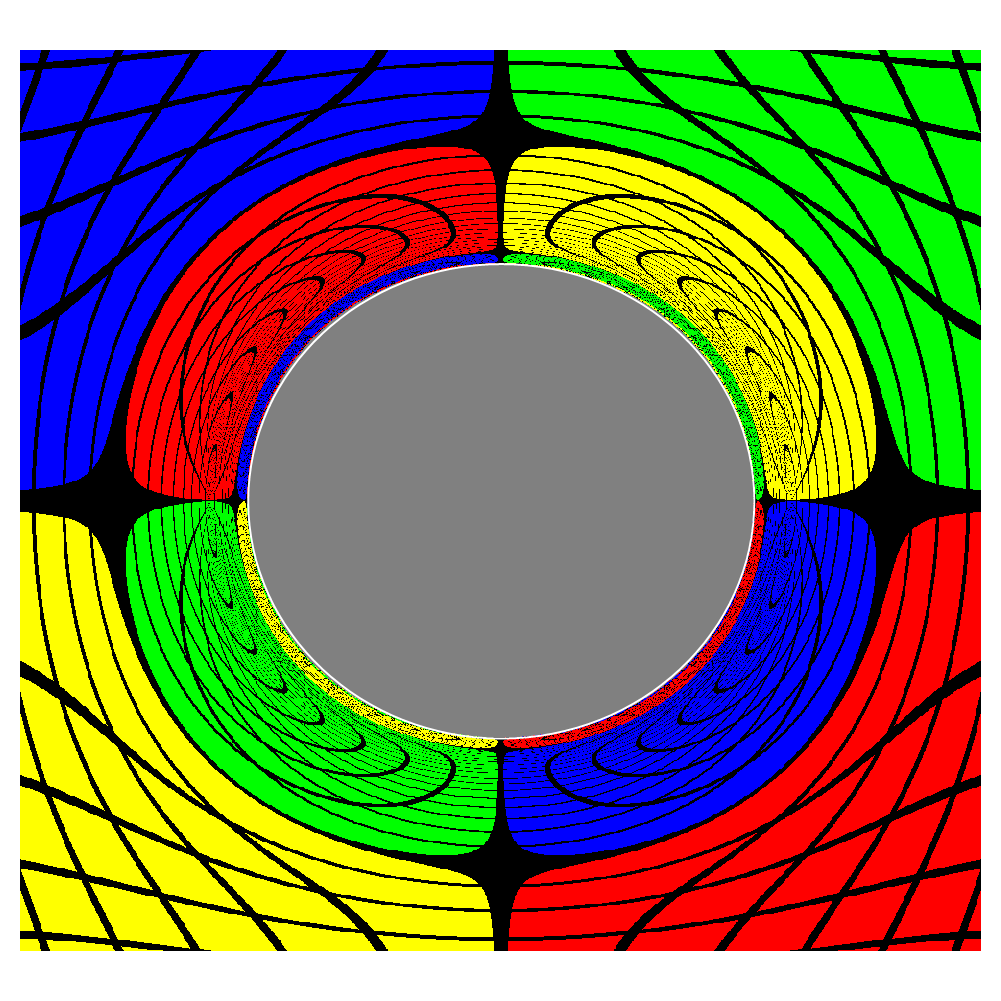}  & \includegraphics[width=0.4\textwidth]{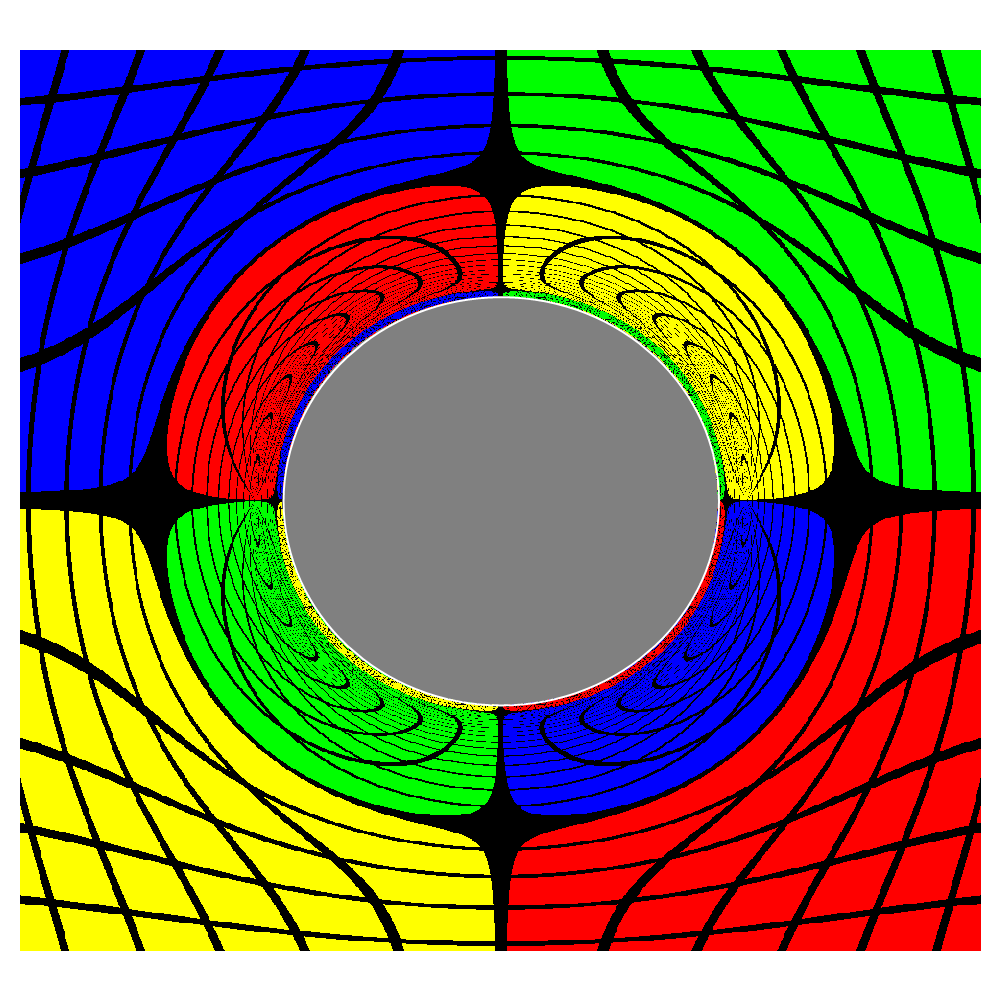}\tabularnewline
\end{tabular}\caption{The images of wormholes illuminated by the celestial sphere on the
same side as the observer. The left panel shows the wormhole with
a single photon sphere $\left(a=5M\right)$, while the right panel
illustrates the case with two photon spheres $\left(a=2.1M\right)$.
The white curve denotes the critical curve, which is generated by
light rays escaping the photon spheres.}
\label{CSatOU} 
\end{figure}

We first investigate the scenario where the observer is situated on
the same side as the celestial sphere (e.g., $r_{\text{cs}}=50M$).
The observed images are presented in FIG. \ref{CSatOU}. When the
impact parameter is smaller than the critical impact parameter, light
rays originating from the celestial sphere will inevitably pass through
the throat into the other spacetime, making them undetectable to the
observer. Consequently, the images exhibit a dark region confined
by the critical curve, which resembles black hole images. Additionally,
the left panel displays a larger dark region due to its larger critical
impact parameter. Moreover, a series of compressed higher-order celestial
sphere images exist outside the critical curve, asymptotically approaching
it. Interestingly, wormholes with two photon spheres exhibit similar
observational appearance to those with one, as the observed photons
are solely influenced by the photon sphere at $r=r_{\text{ph}}^{+}$.

\begin{figure}[ptb]
\begin{tabular}{cc}
\includegraphics[width=0.4\textwidth]{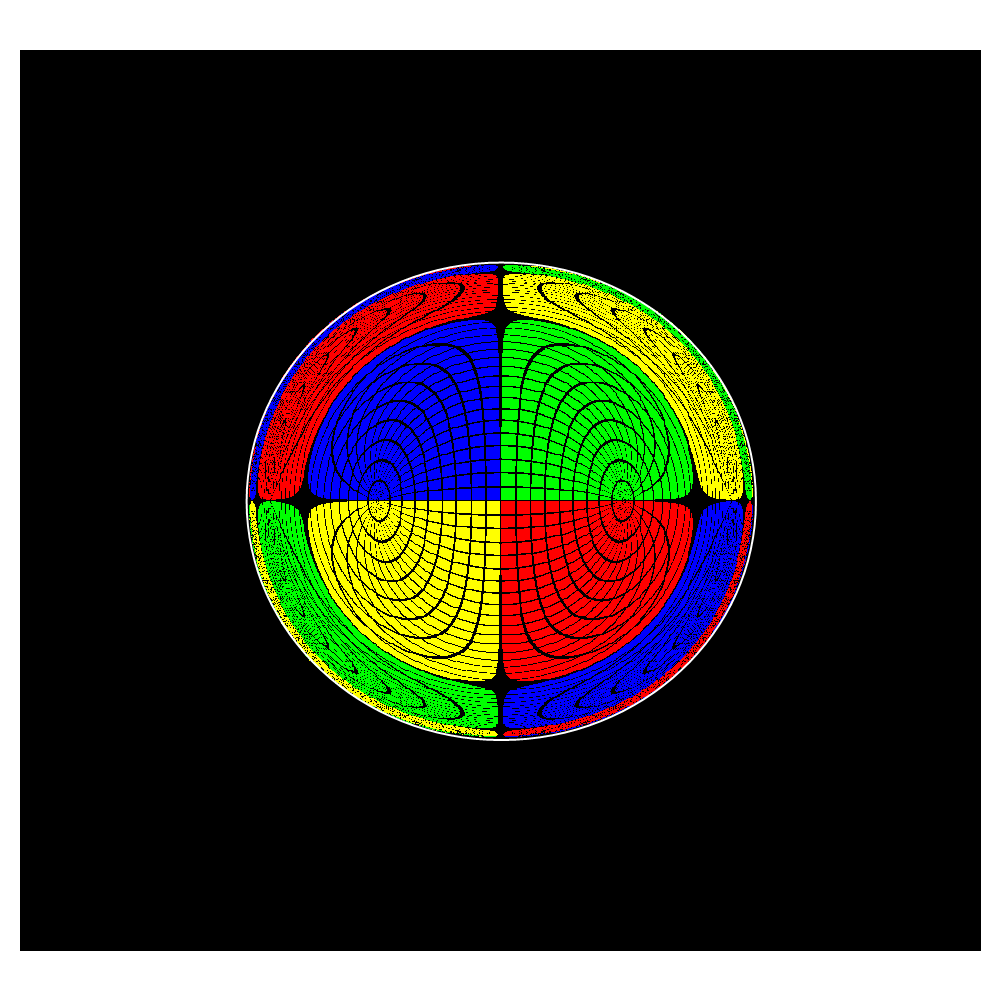}  & \includegraphics[width=0.4\textwidth]{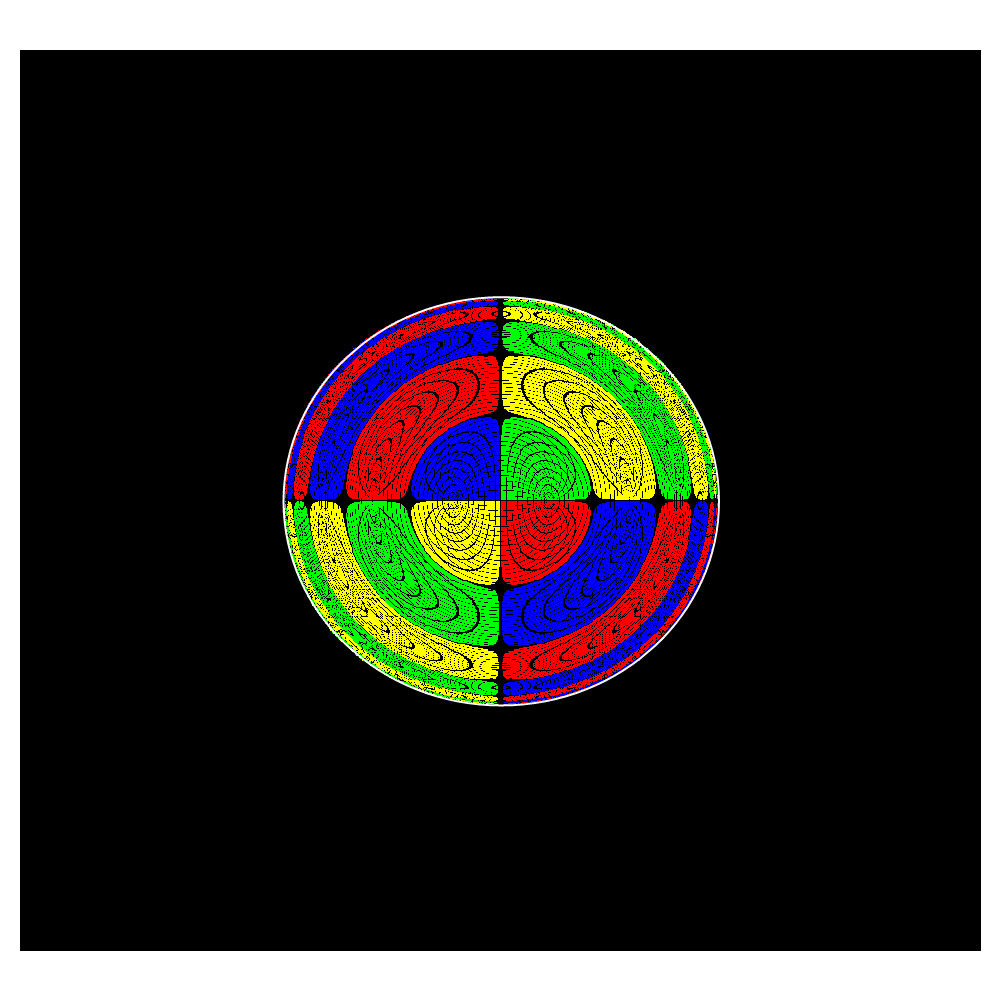}\tabularnewline
\end{tabular}\caption{The images of wormholes illuminated by the celestial sphere on the
opposite side from the observer. The left panel shows the wormhole
with a single photon sphere $\left(a=5M\right)$, while the right
panel illustrates the case with two photon spheres $\left(a=2.1M\right)$.
Since photons emitted from the celestial sphere must traverse the
throat before reaching the observer, the observed images are bounded
by the critical curve. Compared to the single photon sphere case,
the image of the double-photon sphere wormhole exhibits more compressed
higher-order images. The critical curve, outlined in white, is traced
by light rays escaping the photon spheres.}
\label{CSatAU} 
\end{figure}

To investigate the optical appearance of the celestial sphere on the
other side of the wormhole, we extend our analysis to scenarios where
the celestial sphere is located at $r_{\text{cs}}=-50M$. The left
and right panels of FIG. \ref{CSatAU} display the corresponding image
of wormholes with one and two photon spheres, respectively. Photons
emitted from the celestial sphere in the other spacetime can traverse
the throat and be observed, forming celestial sphere images within
the critical curve. Furthermore, the wormhole with two photon spheres
exhibits a greater number of compressed higher-order celestial sphere
images compared to the single-photon sphere case. This phenomenon
occurs because, in the two-photon sphere wormhole, photons emitted
from the celestial sphere traverses both photon spheres at $r=r_{\text{ph}}^{-}$
and $r=r_{\text{ph}}^{+}$ before reaching the observer. As the impact
parameter approaches the critical value, the increased light deflection
due to the double effect of the photon spheres results in more compressed
higher-order celestial sphere images.

Our analysis reveals that the relative positioning between the celestial
sphere and the observer has a significant influence on the observed
image. When the celestial sphere is located on the same side as the
observer, light rays are primarily affected by one photon sphere,
resulting in an image similar to that of a Schwarzschild black hole.
However, when the celestial sphere and the observer are on different
sides, the observed images are within the critical curve. In this
scenario, the presence of two photon spheres leads to more compressed
higher-order celestial sphere images.

\section{Hot Spot}

\label{sec:Hot-Spot}

This section explores observable signatures of hot spots orbiting
around traversable wormholes. To simplify the analysis, we consider
the hot spot as an isotropically emitting sphere with a radius of
$0.25M$, following a counterclockwise path on the ISCO. As found
in \cite{Chen:2023uuy,Chen:2023knf}, observers with large inclination
angles perceive more significant Doppler effects during the hot spot
motion, which can provide more information about spacetime properties.
Therefore, the observers in this section are placed at $\left(r_{o},\varphi_{o},\theta_{o}\right)=\left(100M,\pi,80^{\circ}\right)$.
To achieve optimal precision and efficiency, we simulate the hot spot
image using a $1000\times1000$ pixel grid for each snapshot and generate
$500$ snapshots for a complete orbit. Light rays are traced backward
from the observer to the hot spot at each time $t_{k}$ to determine
the intensity $I_{klm}$ assigned to each pixel. To provide valuable
insights into the characteristics and evolution of hot spot images
within a single orbital period, our analysis focuses on the following
image properties \cite{Hamaus:2008yw,Rosa:2022toh,Rosa:2023qcv,Tamm:2023wvn,Rosa:2024bqv}: 
\begin{itemize}
\item Time-integrated image: This image captures the complete trajectory
of the hot spot in one period by integrating the intensity over all
snapshots. Mathematically, each pixel is assigned an integrated intensity,
\begin{equation}
\left\langle I\right\rangle _{lm}=\sum_{k}I_{klm}.
\end{equation}
\item Light curve: The light curve depicts the variation in the magnitude
over time. The flux at each snapshot is calculated using the following
formula, 
\begin{equation}
F_{k}=\sum\limits _{l}\sum\limits _{m}\Delta\Omega I_{klm}.
\end{equation}
The magnitude is then derived from the flux as follows, 
\begin{equation}
m_{k}=-2.5\lg\left[F_{k}/\min\left(F_{k}\right)\right].
\end{equation}
\item Centroid motion: The centroid motion tracks the movement of the hot
spot image. In each snapshot, the centroid is obtained by calculating
the intensity-weighted average position of all pixels, normalized
by the total flux, 
\begin{equation}
\overrightarrow{c_{k}}=F_{k}^{-1}\sum\limits _{l}\sum\limits _{m}\Delta\Omega I_{klm}\overrightarrow{r_{lm}}.
\end{equation}
Here, $\Delta\Omega$ is the solid angle per pixel, and $\overrightarrow{r_{lm}}$
is the position of the pixel relative to the image center. 
\end{itemize}
\begin{figure}[ptb]
\begin{tabular}{cc}
\includegraphics[width=0.4\textwidth]{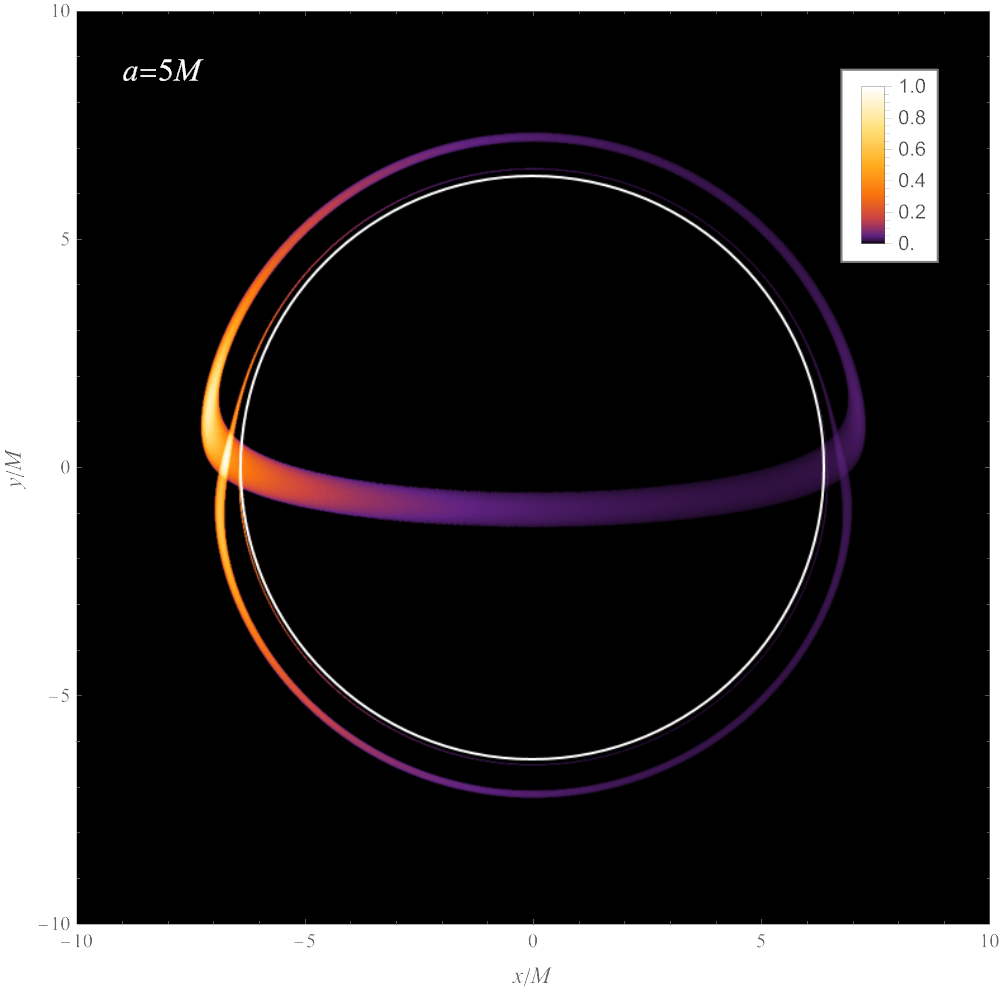}  & \includegraphics[width=0.4\textwidth]{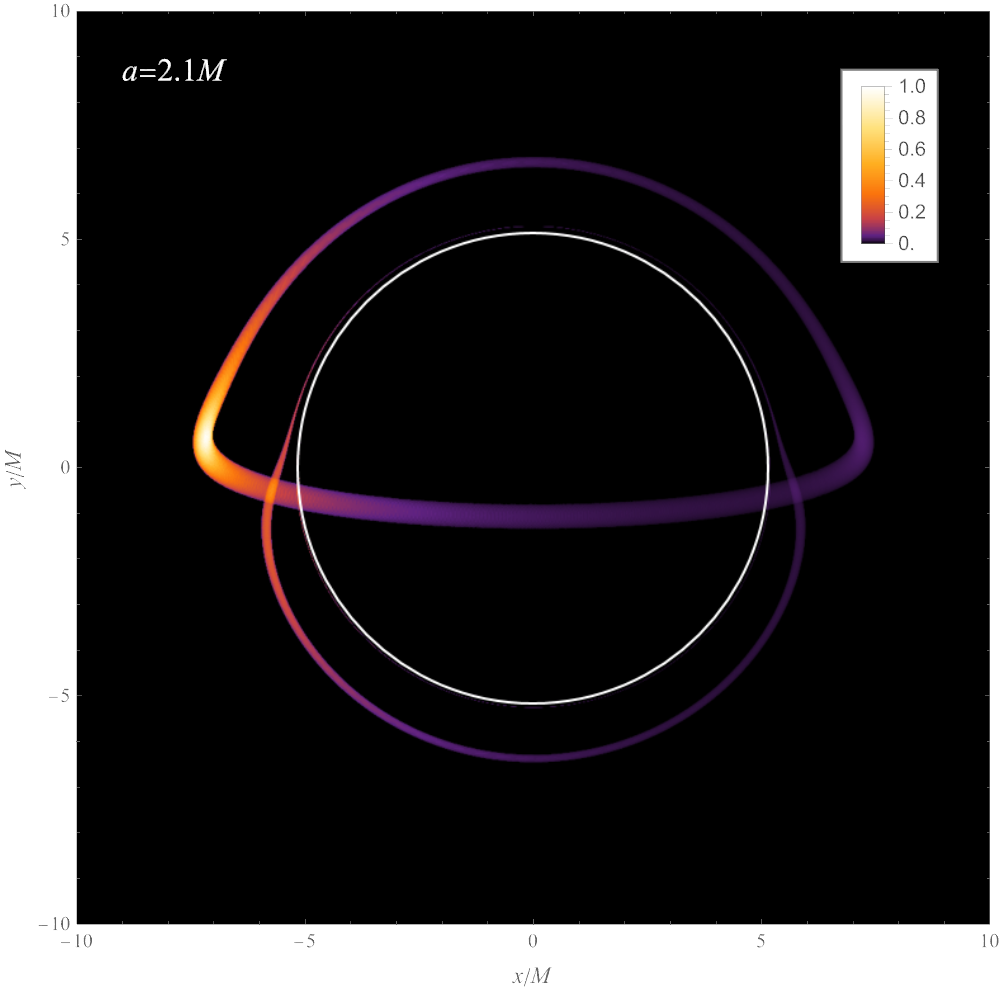}\tabularnewline
\end{tabular}\caption{The time-integrated images of the hot spot orbiting wormholes on the
same side as the observer. The observer is positioned at an inclination
angle of $\theta_{o}=80^{\circ}$. Pixel intensities are normalized
to the maximum value for comparison. The white curve represents the
critical curve, which is generated by light rays with the critical
impact parameter. Both the images of the single-photon sphere (\textbf{Left
Panel}) and double-photon sphere (\textbf{Right Panel}) wormholes
show two distinct image tracks, closely resembling the Schwarzschild
black hole case.}
\label{HSatOU} 
\end{figure}

FIG. \ref{HSatOU} presents time-integrated images of the hot spot
moving on the same side as the observer. Similar to observations in
black holes \cite{Rosa:2022toh,Chen:2024ilc}, both single-photon
sphere and double-photon sphere wormholes exhibit two distinct image
tracks. Additionally, the two image tracks in the left panel are closer
to the critical curve, as the corresponding wormhole with a larger
$a$ has a larger critical curve. To decipher the origin of these
tracks, we introduce a numerical count, $n$, representing the number
of equatorial plane crossings a light ray undergoes during its trajectory.
This numerical count characterizes light rays and the resulting image
tracks. Consequently, the semicircular track corresponds to the primary
image with $n=0$, while the other one corresponds to the secondary
image with $n=1$. Furthermore, when the hot spot travels in front
of the wormhole, its primary image forms the lower section of the
semicircle track, and its secondary image forms the upper section
closer to the critical curve. Conversely, if the hot spot moves behind
the wormhole, the positions of the primary and secondary images are
reversed.

\begin{figure}[ptb]
\includegraphics[width=0.4\textwidth]{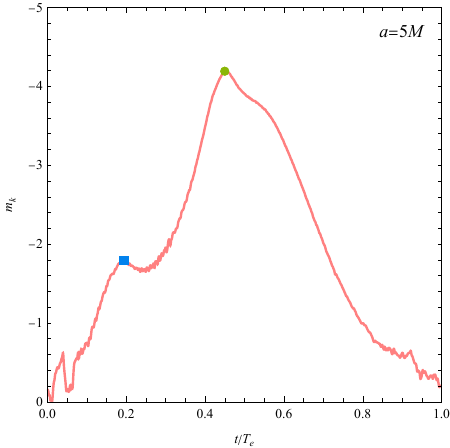}\includegraphics[width=0.4\textwidth]{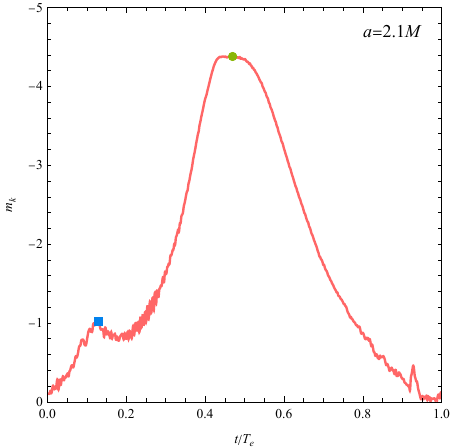}

\includegraphics[width=0.4\textwidth]{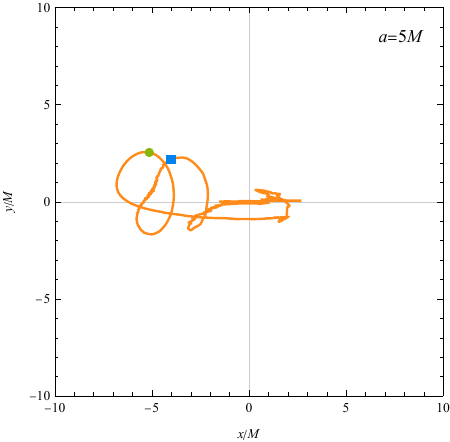}\includegraphics[width=0.4\textwidth]{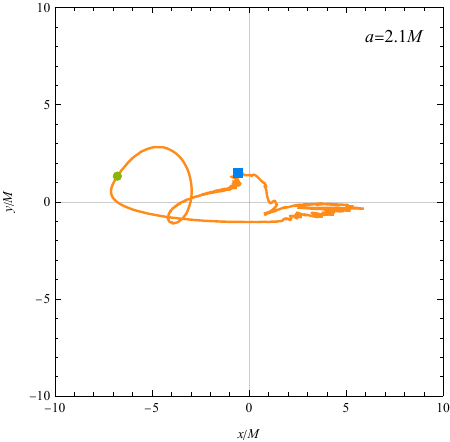}\caption{The light curve (\textbf{Top Row}) and the centroid motion (\textbf{Bottom
Row}) for the hot spot that moves on the same side as the observer.
The left and right columns correspond to the wormholes with a single
and double photon spheres, respectively. Green dots and blue squares
denote the highest and second-highest peaks of the light curve, respectively.}
\label{TMandCatOU} 
\end{figure}

\begin{figure}[ptb]
\begin{centering}
\includegraphics[width=0.4\textwidth]{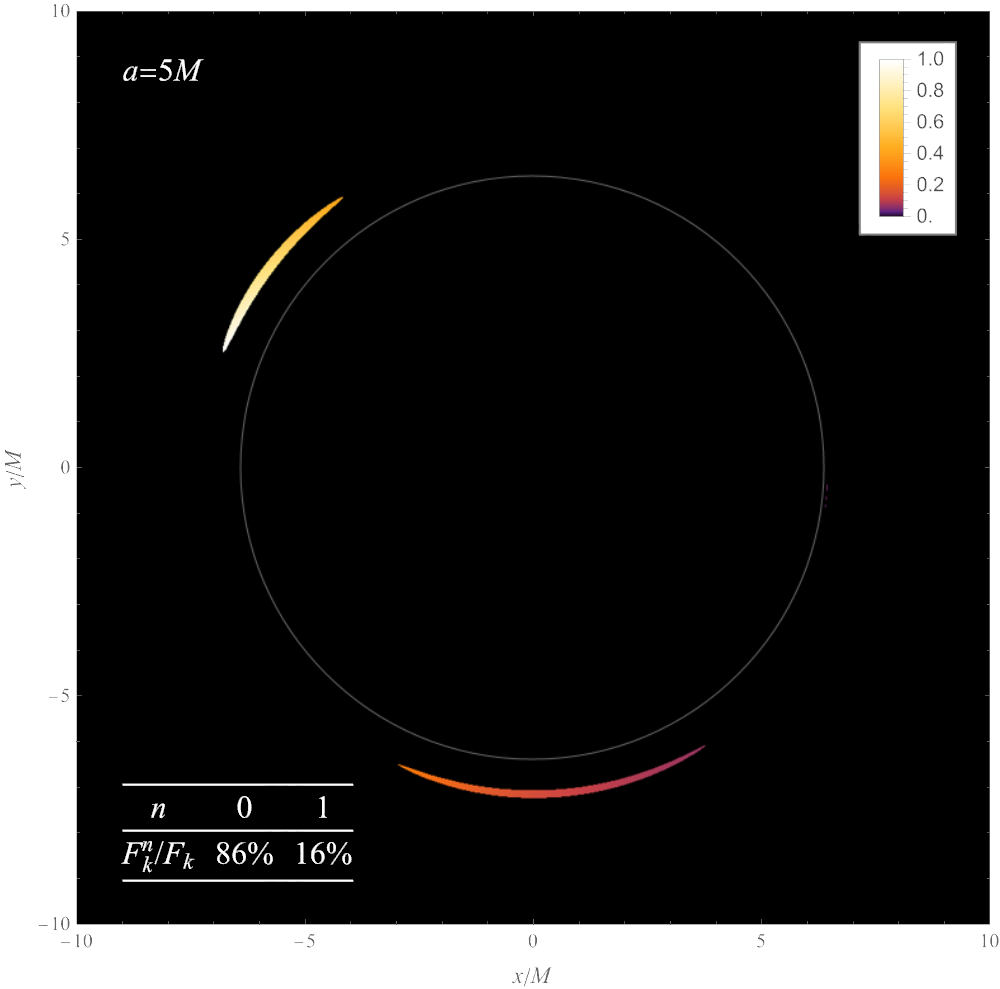}\includegraphics[width=0.4\textwidth]{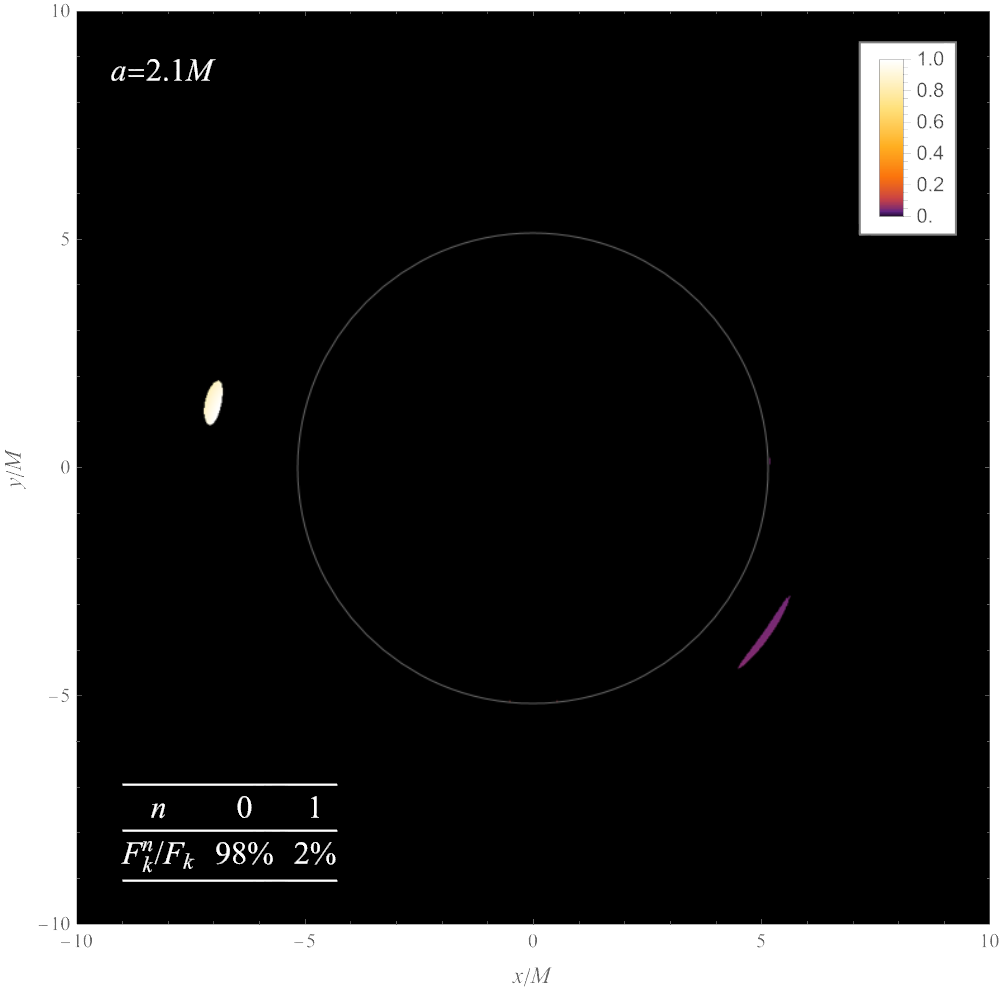} 
\par\end{centering}
\begin{centering}
\includegraphics[width=0.4\textwidth]{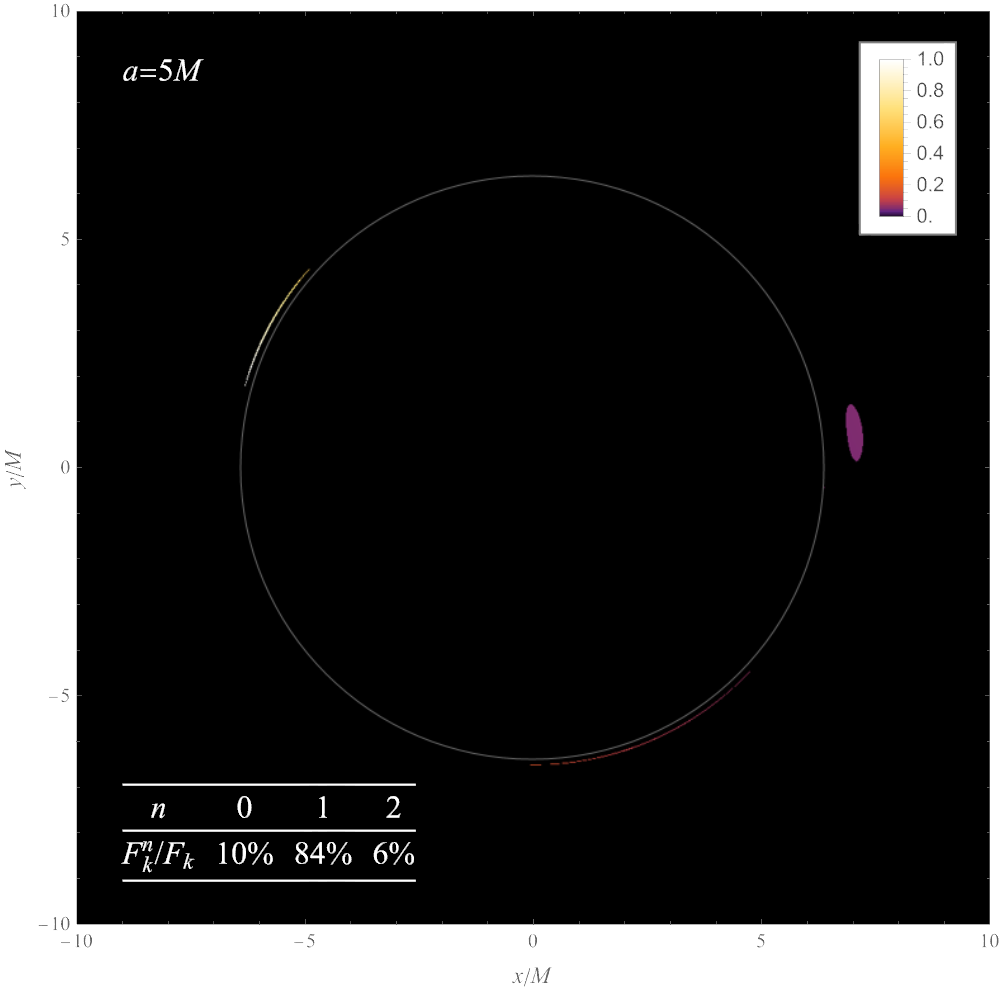}\includegraphics[width=0.4\textwidth]{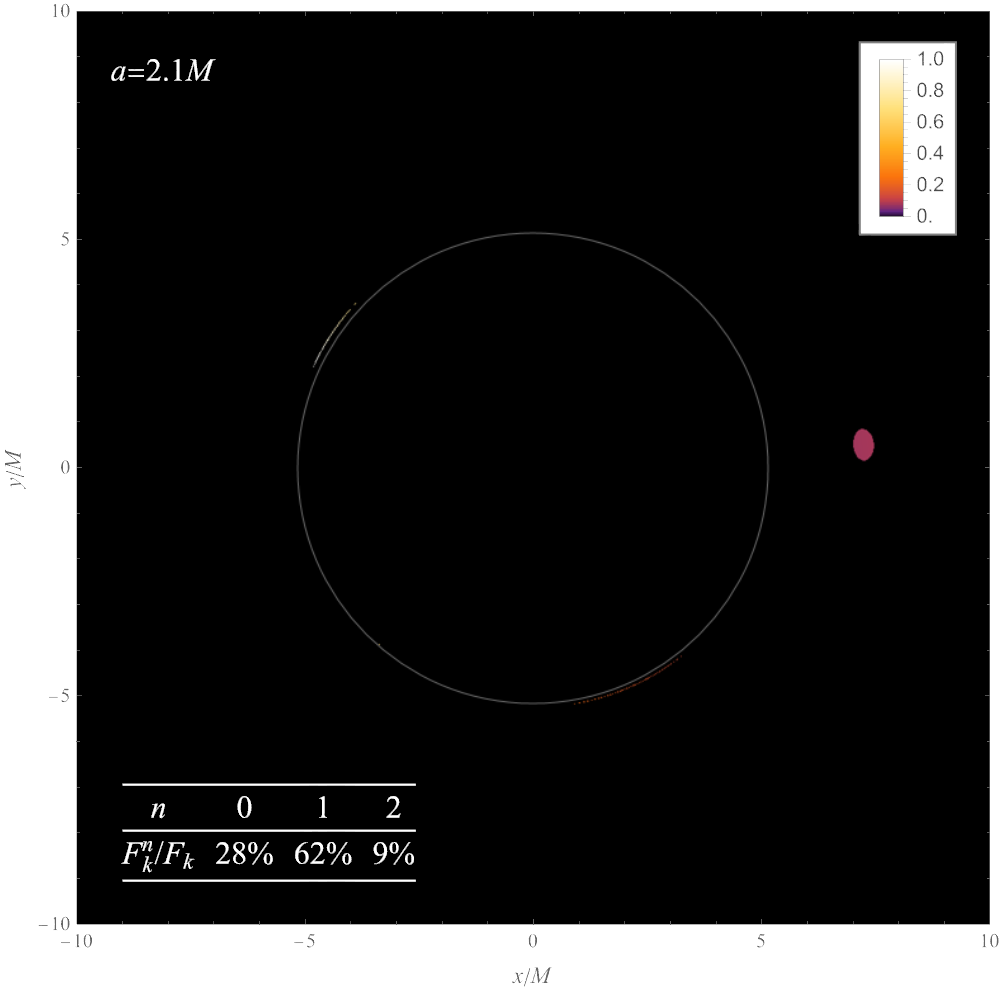} 
\par\end{centering}
\centering{}\caption{Snapshots of the same-side hot spot for the single-photon sphere (\textbf{Left
Column}) and double-photon sphere (\textbf{Right Column}) wormholes,
captured at the light curve peaks. The upper and lower rows depict
the snapshots corresponding to the highest and second-highest peaks,
respectively. The relative contribution of the $n^{\text{th}}$-order
image to the total flux $F_{k}$ is given by $F_{k}^{n}/F_{k}$, where
$F_{k}^{n}$ denotes the flux of the $n^{\text{th}}$-order image
at $t=t_{k}$.}
\label{SatOU} 
\end{figure}

The light curve and centroid motion of the hot spot images are depicted
in the top and bottom rows of FIG. \ref{TMandCatOU}, respectively.
As expected, both single-photon sphere and double-photon sphere wormholes
exhibit light curve and centroid motion characteristics similar to
those observed in Schwarzschild black holes. Specifically, the light
curve displays a prominent peak (denoted by green dots) and a secondary,
fainter peak (denoted by blue squares). The Doppler effect displaces
the centroid towards the left within the field of view, and the centroid
motion appears irregular due to the presence of higher-order images.
These findings suggest that, when hot spots and observers are on the
same side, the appearance of traversable wormholes can mimic that
of Schwarzschild black holes. Additionally, the snapshot of the highest
and second-highest light curve peaks are presented in FIG. \ref{SatOU}.
These figures reveal that the primary $\left(n=0\right)$ and secondary
$\left(n=1\right)$ images dominate the highest and second-highest
peaks, respectively.

\begin{figure}[ptb]
\begin{tabular}{cc}
\includegraphics[width=0.4\textwidth]{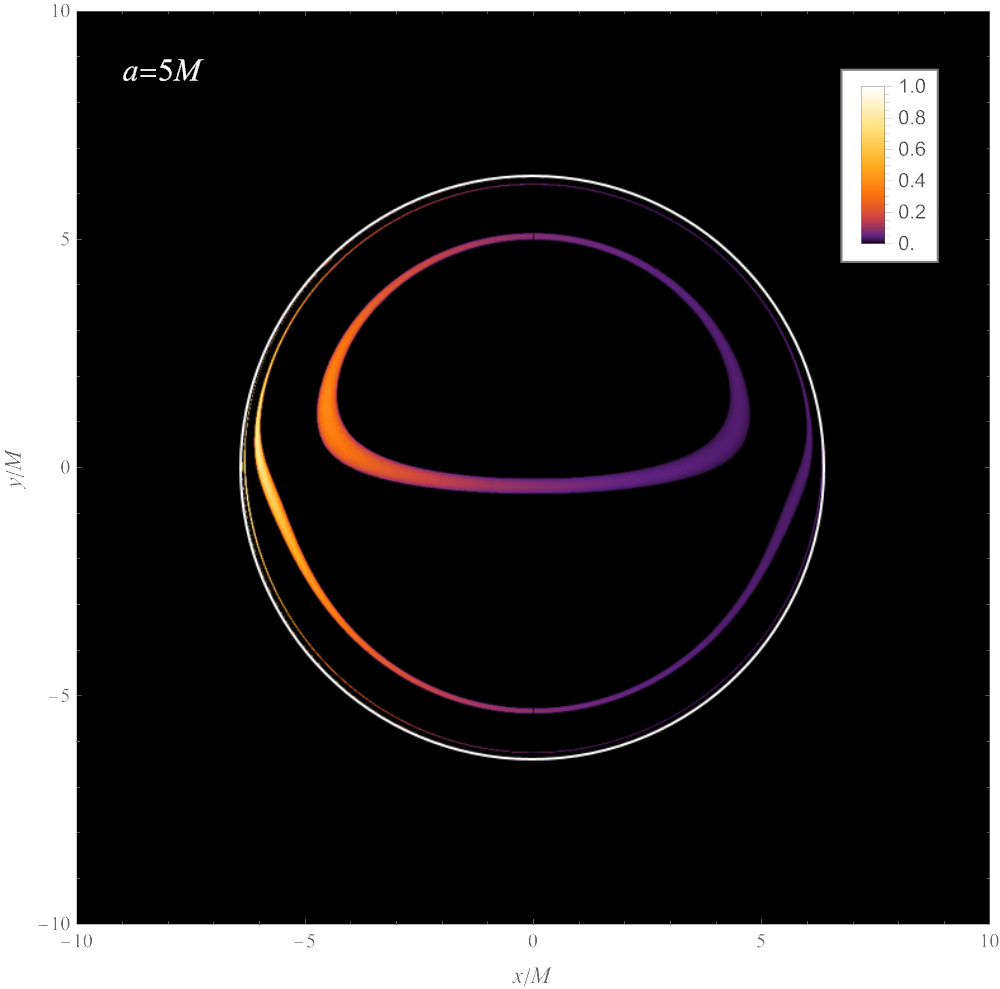}  & \includegraphics[width=0.4\textwidth]{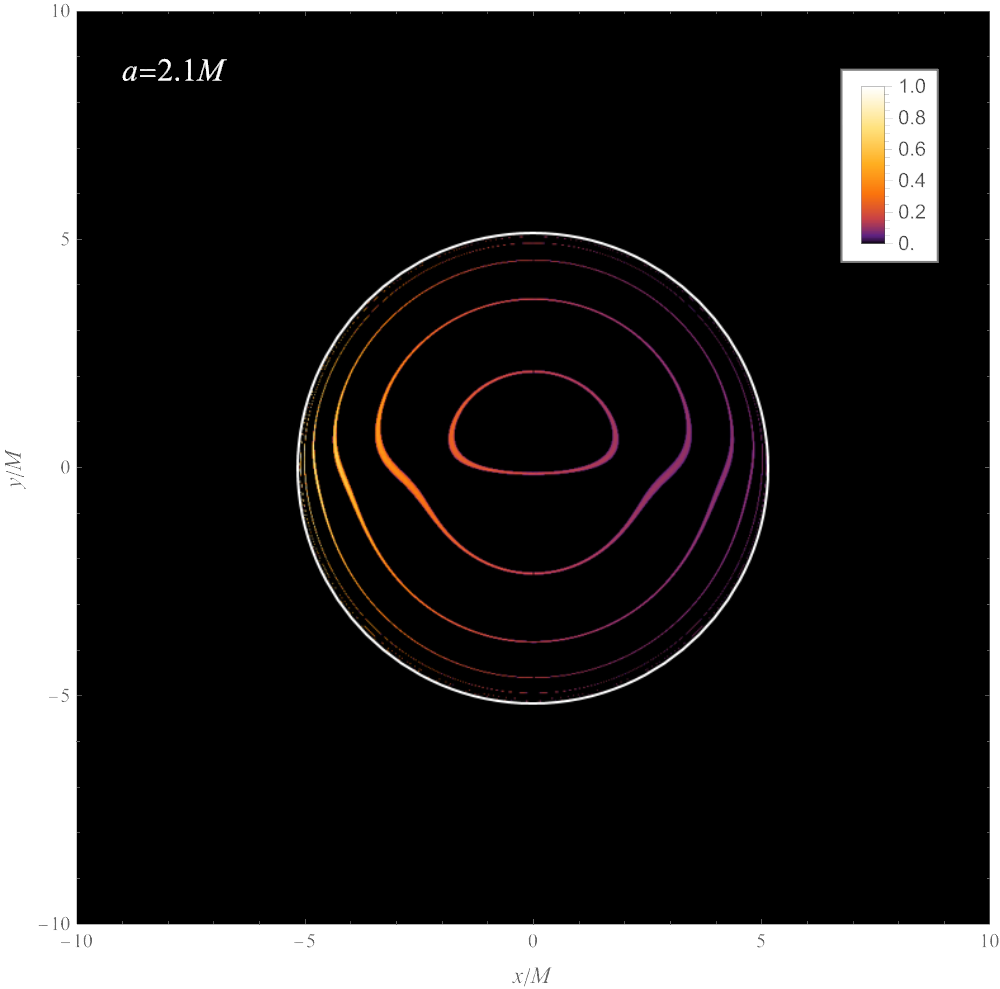}\tabularnewline
\end{tabular}\caption{The time-integrated images of the hot spot orbiting wormholes on the
side opposite to the observer. The critical curve is depicted as white
curves. \textbf{Left Panel}: The image of the single-photon sphere
wormhole exhibits two distinct image tracks within the critical curve.
\textbf{Right Panel}: The image of the double-photon sphere wormhole
shows four distinct image tracks within the critical curve. Traversing
an additional photon sphere results in the emergence of two more image
tracks.}
\label{HSatAU} 
\end{figure}

We now consider the scenario in which the hot spot is located on the
opposite side of the observer. Specifically, the hot spot orbits the
wormholes along the ISCO at $r=r_{e}^{-}$. Time-integrated images
for this scenario are presented in FIG. \ref{HSatAU}. Only photons
with impact parameters less than the critical value, $b<b_{\text{ph}}$,
can overcome the potential peak(s) and traverse the wormhole. Consequently,
the images are confined within the critical curve. For the single-photon
sphere wormhole, the hot spot image exhibits two distinct tracks.
The inner and outer tracks correspond to the $n=1$ and $n=2$ images,
respectively. Note that photons traversing the wormhole throat always
cross the equatorial plane once, resulting in the absence of the $n=0$
image. On the other hand, the double-photon sphere wormhole presents
a markedly different observation image compared to the single-photon
case. Since emitted photons need to traverse two photon spheres to
reach the observer, stronger light deflections occur, which manifests
as four distinct tracks in the integrated image. From the innermost
outward, these tracks correspond to images with $n=1$, $2$, $3$
and $4$, respectively.

\begin{figure}[ptb]
\includegraphics[width=0.4\textwidth]{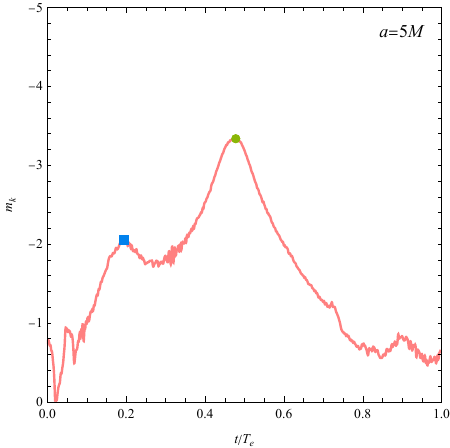}\includegraphics[width=0.4\textwidth]{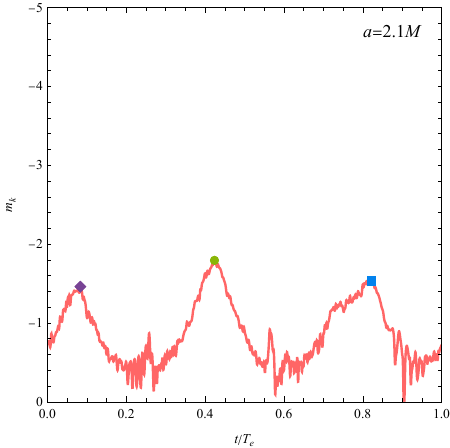}

\includegraphics[width=0.4\textwidth]{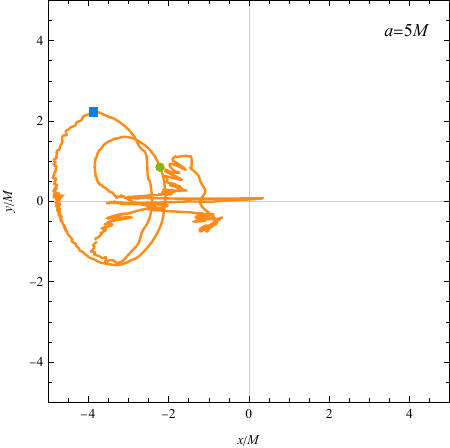}\includegraphics[width=0.4\textwidth]{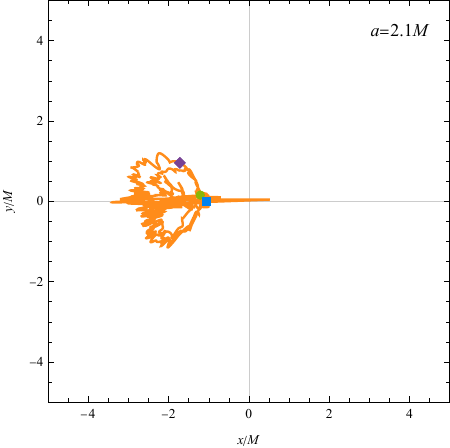}\caption{The light curve (\textbf{Top Row}) and the centroid motion (\textbf{Bottom
Row}) for hot spots orbiting wormholes on the side opposite to the
observer. The left and right columns correspond to the wormholes with
a single and double photon spheres, respectively. The light curve
in the single-photon sphere case exhibit two peaks, while three peaks
of comparable heights appear in the double-photon sphere case.}
\label{TMandCatAU} 
\end{figure}

\begin{figure}[ptb]
\begin{centering}
\begin{tabular}{cc}
\includegraphics[width=0.4\textwidth]{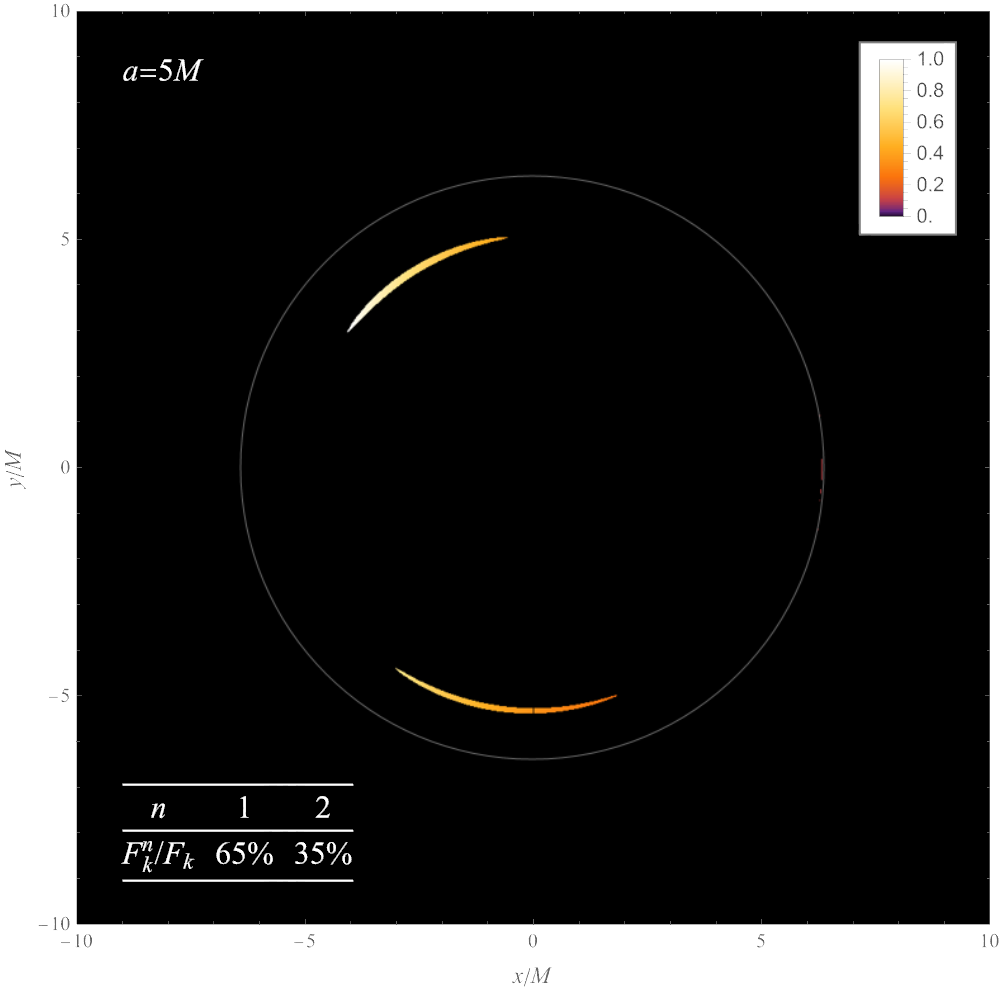}  & \includegraphics[width=0.4\textwidth]{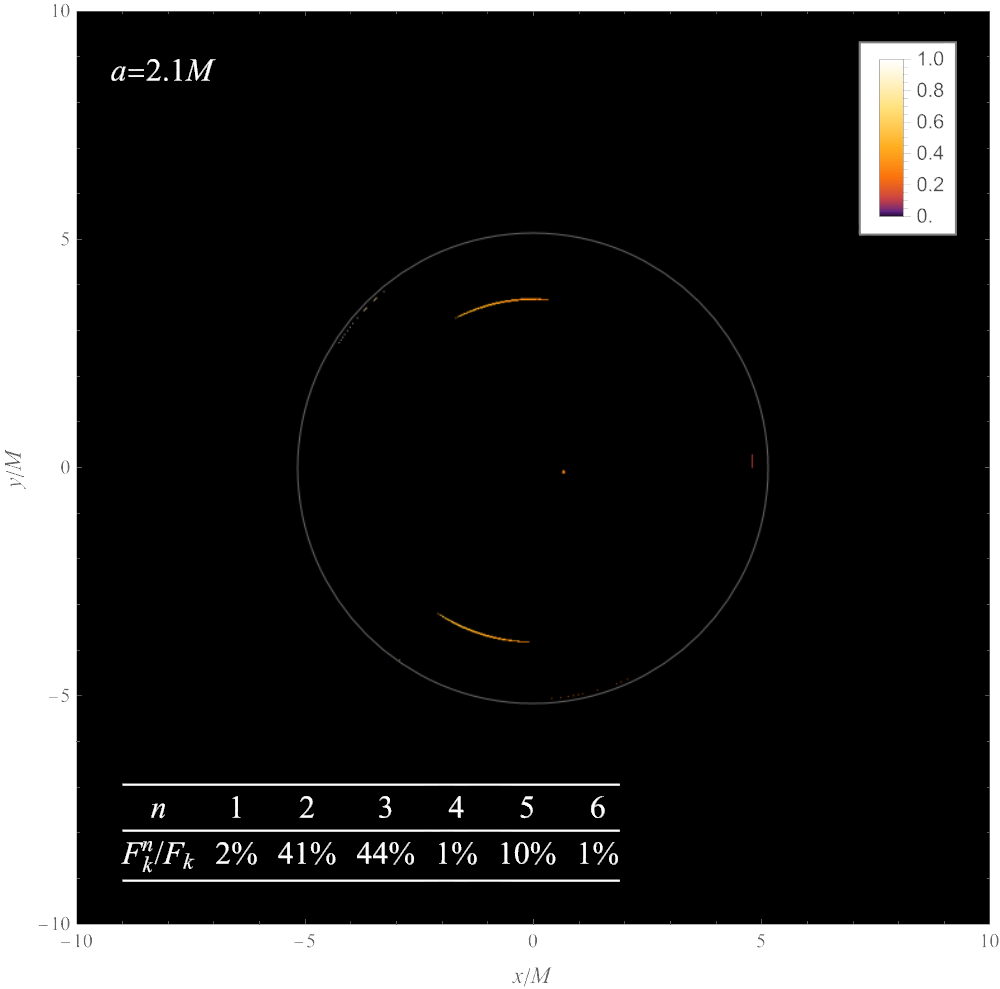}\tabularnewline
\includegraphics[width=0.4\textwidth]{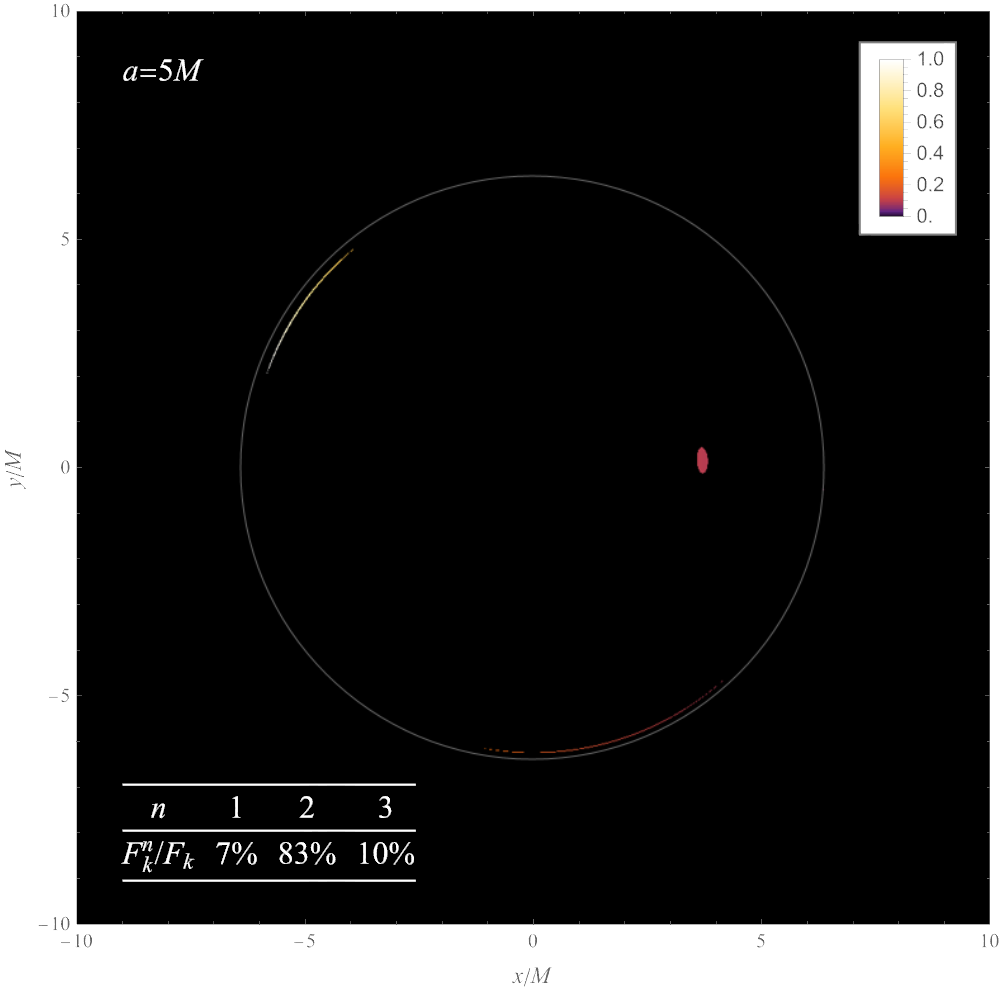}  & \includegraphics[width=0.4\textwidth]{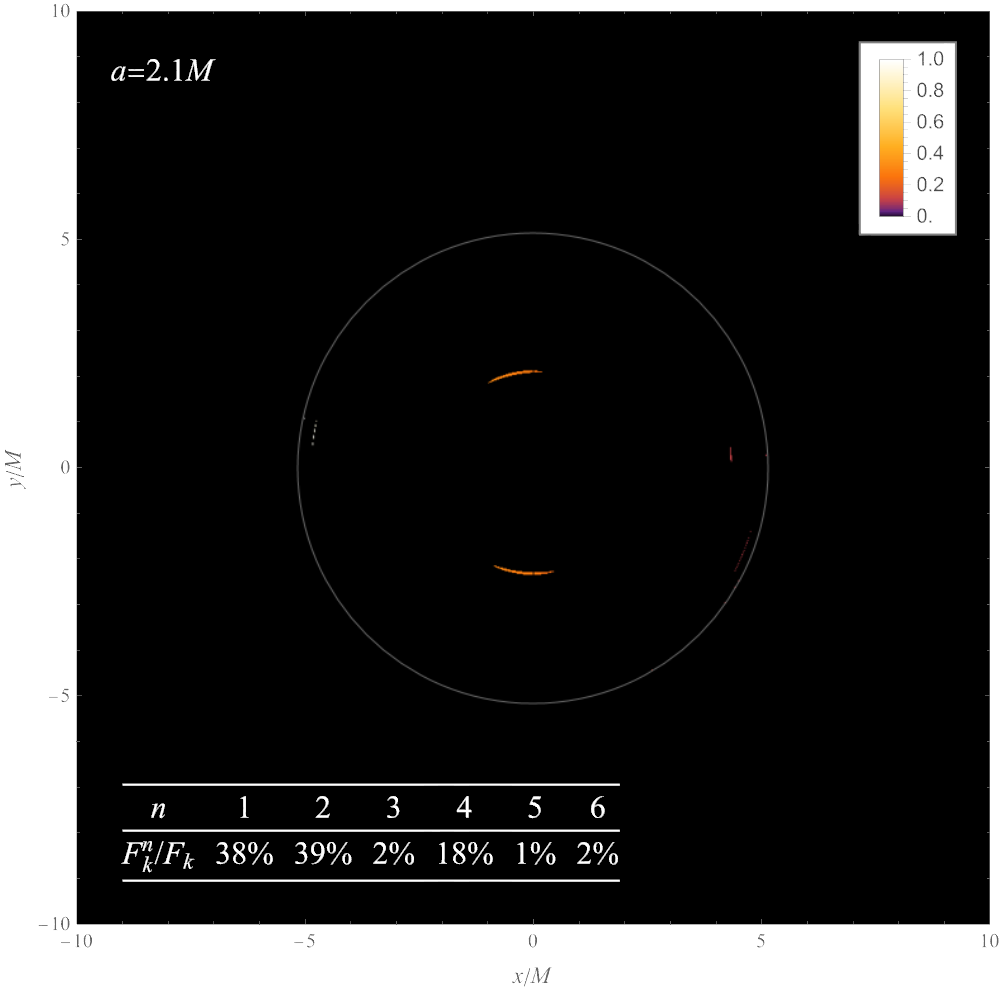}\tabularnewline
 & \includegraphics[width=0.4\textwidth]{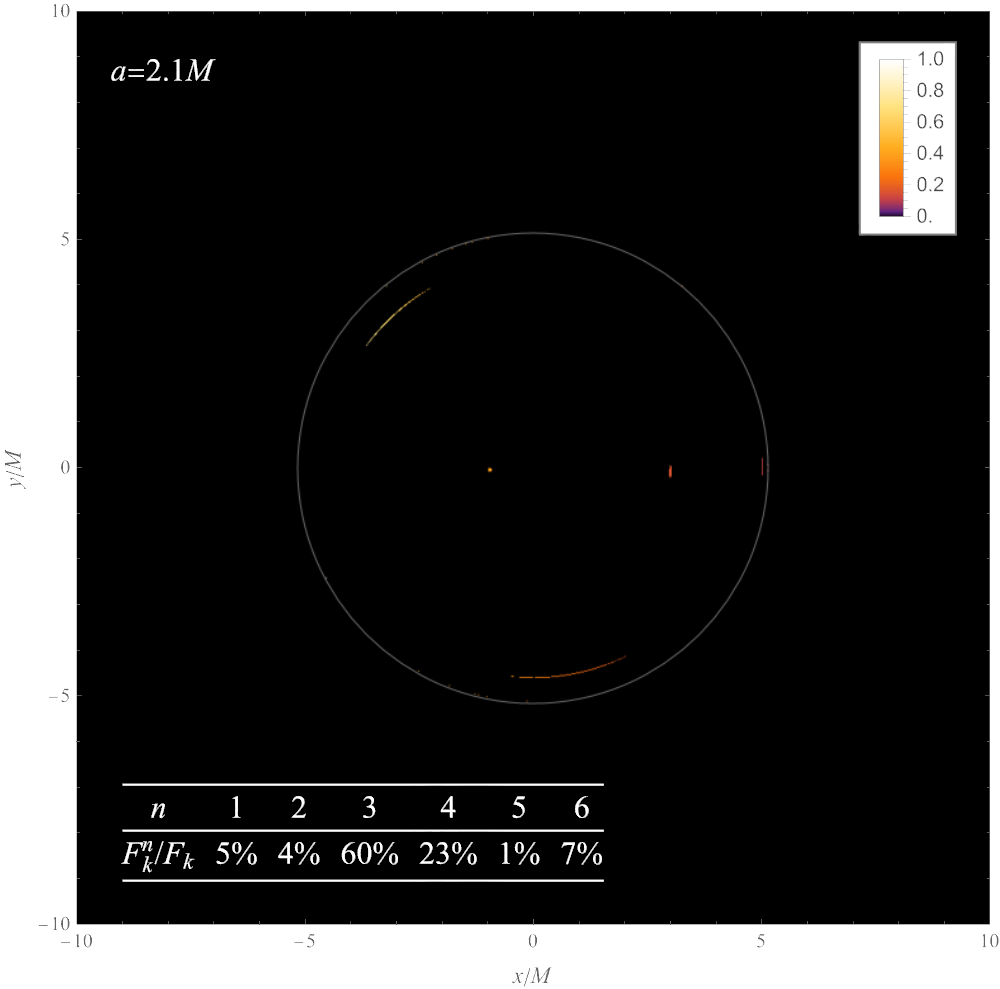}\tabularnewline
\end{tabular}
\par\end{centering}
\caption{Snapshots of the hot spot orbiting wormholes on the side opposite
to the observer, for wormholes with a single (\textbf{Left Column})
and double (\textbf{Right Column}) photon spheres. The snapshots are
captured at the highest (\textbf{Top Row}), second-highest (\textbf{Middle
Row}) and third-highest (\textbf{Bottom Row}) magnitude peaks.}
\label{SatAU} 
\end{figure}

The corresponding light curve and centroid motion of the hot spot
are presented in the top and bottom rows of FIG. \ref{TMandCatAU},
respectively. Similar to the same-side case, the single-photon sphere
wormhole exhibits a light curve with a prominent peak and a secondary
peak. Consistent with this, the snapshots in the left panel of FIG.
\ref{SatAU} reveals that the highest and second-highest peaks are
dominated by the $n=1$ and $n=2$ images, respectively. Furthermore,
due to the confinement of the images within the critical curve, the
centroid motion is restricted to a smaller range. Intriguingly, the
hot spot orbiting the double-photon sphere wormhole displays three
distinct peaks of comparable heights in the light curve, as shown
in the upper-right panel of FIG. \ref{TMandCatAU}. The corresponding
snapshots in the right panels of FIG. \ref{SatAU} reveal that the
highest peak is dominated by a combination of $n=2$ and $3$ images,
the second-highest peak by $n=1$ and $2$ images, and the third-highest
peak by $n=3$ and $4$ images. Consequently, the presence of more
higher-order images leads to a more irregular centroid motion in this
case.

These observations of the different-side hot spot offer potential
avenues for distinguishing wormholes from black holes. Notably, wormholes
exhibit a more restricted range of central motion compared to black
holes. Additionally, the three-peaked light curve of the double-photon
sphere wormhole provides a unique signature for differentiating it
from the single-photon sphere counterpart.

\section{Conclusions}

\label{sec:Conclusions}

This paper investigates the images of traversable wormholes illuminated
by celestial spheres and hot spots. Two spacetimes are connected at
the wormhole throat, which allows light rays to travel from one side
to the other one. When the celestial sphere and hot spot reside on
the same side as the observer, their images closely resemble those
of a black hole. However, significant differences arise when the light
sources are positioned on the different side from the observer. Specifically,
we found that 
\begin{itemize}
\item Same-side scenario: The celestial sphere forms images outside the
critical curve, with a sequence of compressed celestial sphere images
asymptotically approaching the critical curve. The hot spot exhibits
two distinct image tracks with asymmetric brightness when observed
at an inclination angle of $\theta_{o}=80^{\circ}$. These two tracks
lead to two peaks in the light curve. 
\item Different-side scenario: The celestial sphere always forms images
within the critical curve, with compressed higher-order images asymptotically
approaching the critical curve. The hot spot manifests as two distinct
image tracks for the single-photon sphere case, while the double-photon
sphere case exhibits four distinct image tracks. In the single-photon
sphere case, the two tracks again lead to two peaks in the light curve.
Conversely, the four tracks in the double-photon case result in three
peaks in the light curve. 
\end{itemize}
By analyzing these image characteristics, we can gain valuable insights
into the optical signatures of light sources near wormholes. This
understanding has the potential to not only differentiate between
wormholes and black holes but also discriminate between single-photon
sphere and double-photon sphere wormholes. While observing images
from the other side of the wormhole presents an intriguing prospect,
it also poses greater challenges. Therefore, higher resolution instruments,
such as the next-generation Very Long Baseline Interferometry, are
also keenly expected. 
\begin{acknowledgments}
We are grateful to Tianshu Wu for useful discussions and valuable
comments. This work is supported in part by NSFC (Grant No. 12105191,
12275183, 12275184 and 11875196). 
\end{acknowledgments}

 \bibliographystyle{unsrturl}
\bibliography{ref}

\end{document}